# Temporal Transfer Matrix Method for Exceptional-Point Media via Canonical Basis Expansion

Neng Wang and Guo Ping Wang*

*State Key Laboratory of Radio Frequency Heterogeneous Integration, College of Physics and Optoelectronic Engineering, Shenzhen University, Shenzhen*

*518060, China*

Corresponding to*: gpwang@szu.edu.cn

We present a generalized temporal transfer matrix method (TTMM) for time-varying media that accurately captures wave dynamics in media operating at exceptional points (EPs). The method expands wave fields in the canonical basis of each temporal layer and derives the complete time evolution of all basis vectors. Temporal matching and phase-delay matrices are constructed from the generalized modal matrices and their corresponding eigenvalues. Additionally, an amplitude-boosting matrix is introduced to account for the power-law amplification of field amplitudes associated with EP dynamics. This matrix depends only on the order of the EP and naturally reduces to the identity matrix in its absence. The proposed TTMM is validated through two representative EP media, demonstrating its accuracy and broad applicability.

## I. Introduction

The temporal transfer matrix method (TTMM) has proven to be a powerful framework for studying time-varying systems [1-5], particularly in analyzing temporal multilayer structures [6-8] and photonic time crystals (PTCs) with periodic [9-11] or quasi-periodic [12-14] sharp time interfaces. Early applications of the TTMM primarily focused on isotropic and anisotropic dielectric media while neglecting material dispersion [15-17]. Only recently has the method been extended to dispersive materials [18]. Incorporating dispersion not only enhances physical realism [19, 20] but also unveils a range of previously overlooked phenomena, such as broadened momentum bandgaps [21, 22], uniform-loss-induced exceptional points (EPs) [23], and amplification of static field amplitudes [24]. However, most existing studies have concentrated on ordinary media whose band dispersions remain far from singularities, leaving systems with extreme parameters largely unexplored. In conventional TTMM formulations, the transfer matrices are constructed from two key components: the temporal matching matrix and the phase-delay matrix. The former arises from enforcing temporal boundary conditions between two complete basis sets, while the latter describes the exponential time evolution of fields within each temporal layer.

EPs are singularities where both the eigenvalues and the corresponding eigenvectors of a system's Hamiltonian coalesce [25]. The nonlinear eigenvalue splitting near EPs greatly enhances the system's sensitivity to perturbations [26-28] and gives rise to a variety of fascinating physical phenomena [29-32]. In ordinary systems, each eigenvector evolves independently with a simple exponential behavior governed by its eigenvalue. At an EP, however, this independence breaks down—the eigenvectors coalesce and no longer form a complete basis. Consequently, the time evolution of the wave states incorporates additional polynomial terms that alter the standard exponential dynamics. Therefore, the conventional TTMM, which considers only temporal matching and phase-delay matrices, cannot fully describe the true time evolution of the wave fields.

In this paper, we develop a generalized TTMM by expanding the wave fields in the canonical basis of each temporal layer and deriving the time evolution of all basis vectors. For media operating at EPs, the basis set includes both the ordinary and generalized eigenvectors of the system. The generalized modal matrices—whose columns are composed of these basis vectors—are then employed to construct the temporal matching matrices. Beyond the conventional phase-delay matrix, we introduce an additional term called the amplitude-boosting matrix, which captures the power-law amplification of field amplitudes associated with EP dynamics. This matrix depends solely on the order of the EP and becomes the identity matrix in the absence of an EP, ensuring that our generalized TTMM naturally reduces to the conventional formulation in ordinary media. We apply this TTMM to compute the time-dependent wave fields and Floquet band structures of two representative EP media: lossless Drude media exhibiting an EP of order 2 (EP-2) and lossy Lorentz dispersive media exhibiting an EP of order 4 (EP-4). The results confirm the accuracy and validity of our generalized TTMM. This approach provides a robust framework for exploring the rich and unconventional physics in time-varying media operating at EPs.

## II. Methodology

Consider that the j-th temporal layer is governed by the following Schrodinger-like equation:

$$i\partial_t \vec{\psi} = \hat{H}_j \cdot \vec{\psi}, (1)$$

where $\hat{H}_j$ and $\vec{\psi}$ are a $n \times n$ matrix and a $n$-dimensional vector, representing the Hamiltonian of this temporal layer and wavefunction, respectively. To solve Eq. (1), we first impose a similarity transformation such that $\hat{H}_j$ is transformed to the Jordan normal form $\hat{J}_j$:

$$i\partial_t(\hat{M}_j^{-1} \cdot \vec{\psi}) = i\partial_t \vec{\psi}' = \hat{M}_j^{-1} \cdot \hat{H}_j \cdot \hat{M}_j \cdot \hat{M}_j^{-1} \cdot \vec{\psi} = \hat{J}_j \cdot \vec{\psi}', (2)$$

where $\hat{M}_j$ is a generalized modal matrix of $\hat{H}_j$, formed by the canonical basis of $\hat{H}_j$, and $\vec{\psi}'$ is the wavefunction in the canonical basis.

### A. Diagonalizable Hamiltonian

If $\hat{H}_j$ is diagonalizable, there are $n$ linearly independent eigenvectors $\vec{\psi}_l^{(j)}$ (though not necessarily orthogonal) which construct a complete basis, referred to as the canonical basis of $\hat{H}_j$. The corresponding modal matrix is then given by:

$$\hat{M}_j = (\vec{\psi}_1^{(j)}, \vec{\psi}_2^{(j)}, ..., \vec{\psi}_n^{(j)}). (3)$$

This leads the Jordan normal form $\hat{J}_j$ to be diagonalized:

$$\hat{J}_j = \text{diag}\{\omega_1^{(j)}, \omega_2^{(j)}, ..., \omega_n^{(j)}\}, (4)$$

where $\omega_l^{(j)}$ represent the eigenvalues of $\hat{H}_j$, corresponding to the eigenvector $\vec{\psi}_l^{(j)}$.

Substituting Eq. (4) into Eq. (2) yields $n$ linearly independent differential equations, from which the expression of $\vec{\psi}'$ is obtained easily. Using Eq. (3) and the transformation $\vec{\psi}' = \hat{M}_j^{-1} \cdot \vec{\psi}$, the wavefunction in the original basis is

$$\vec{\psi}(t) = \sum_{l=1}^{n} c_l^{(j)} \vec{\psi}_l^{(j)} e^{-i\omega_l^{(j)}t}, (5)$$

where $c_l^{(j)}$ are the expansion coefficients in the canonical basis and $t=0$ defines the beginning of j-th temporal layer. If we denote $\vec{\beta}_j = (c_1^{(j)}, c_2^{(j)}, ..., c_n^{(j)})^T$, Eq. (5) yields

$$\vec{\psi}(t) = \hat{M}_j \cdot \hat{D}_j(t) \cdot \vec{\beta}_j, (6)$$

where $\hat{D}_j(t)$ is the socalled temporal phase-delay matrix within the j-th temporal layer:

$$\hat{D}_j(t) = \text{diag}\{e^{-i\omega_1^{(j)}t}, e^{-i\omega_2^{(j)}t}, ..., e^{-i\omega_n^{(j)}t}\}. (7)$$

At the time interface $t = t_j$ formed between the j-th and (j+1)-th layers, assuming that $\hat{H}_{j+1}$ is also diagonalizable, using Eq. (6), the continuity of the wavefunction across the time interface yields:

$$\vec{\psi}(t_j^-) = \hat{M}_j \cdot \hat{D}_j(t_j) \cdot \vec{\beta}_j = \vec{\psi}(t_j^+) = \hat{M}_{j+1} \cdot \vec{\beta}_{j+1}, (8)$$

which leads to the temporal matching matrix:

$$\hat{M}_{j,j+1} = \hat{M}_{j+1}^{-1} \cdot \hat{M}_j. (9)$$

Therefore, Eqs. (7) and (9) define the temporal phase-delay and matching matrices for diagonalizable Hamiltonians, where the operated vectors $\vec{\beta}_j$ serve as the expansion coefficients of the wavefunction in the canonical basis at the beginning of the j-th temporal layers.

## B. Defective Hamiltonian

The situation becomes intricate when $\hat{H}_j$ is defective, where the eigenvalues and their associated eigenvectors coalesce, preventing the eigenvectors from forming a complete basis. This condition is also well-known as the EP. In this case, the canonical basis is constructed by the $n$ linearly-independent generalized eigenvectors of $\hat{H}_j$. Similarly, we can expand the wavefunctions in the canonical basis and derive the transfer matrices. To the best of our knowledge, the TTMM in this case has not been reported yet.

Without loss of generality, we assume that there is an EP of order $m$ (EP-$m$) at $\omega_1^{(j)}$. Then, there are $n-m+1$ ordinary eigenvectors. The generalized eigenvectors of rank $p$ corresponding to $\omega_1^{(j)}$ are obtained according to the Jordan chain [33]:

$$\left(\hat{H}_j - \omega_1^{(j)}\hat{I}_n\right)^p \vec{\psi}_{1,p}^{(j)} = 0, \quad (10)$$

where $\hat{I}_n$ is the $n \times n$ identity matrix, and $p$ runs from 1 to $m$. The generalized eigenvector of rank 1 is the ordinary eigenvector, namely $\vec{\psi}_{1,1}^{(j)} = \vec{\psi}_1^{(j)}$. The generalized modal matrix of $\hat{H}_j$ is given by

$$\hat{M}_j = (\vec{\psi}_{1,m}^{(j)}, \vec{\psi}_{1,m-1}^{(j)}, ..., \vec{\psi}_{1,1}^{(j)}, \vec{\psi}_2^{(j)}, ..., \vec{\psi}_{n-m+1}^{(j)}), \quad (11)$$

and the Jordan normal form of $\hat{H}_j$ is expressed as:

$$\hat{J}_j = \hat{J}_{\omega_1^{(j)},m} \oplus \mathrm{diag}\{\omega_2^{(j)}, \omega_3^{(j)}, ..., \omega_n^{(j)}\}, \quad (12)$$

where $\hat{J}_{\omega_1^{(j)},m}$ denotes a $m \times m$ Jordan block with the eigenvalue $\omega_1^{(j)}$:

$$\hat{J}_{\omega_1^{(j)},m} = \begin{pmatrix} \omega_1^{(j)} & 1 & & \\ & \omega_1^{(j)} & \ddots & \\ & & \ddots & 1 \\ & & & \omega_1^{(j)} \end{pmatrix}. \quad (13)$$

Substituting Eq. (13) into Eq. (2), and using an iterative approach, $\vec{\psi}'$ is solved:

$$\vec{\psi}' = (X_1, X_2, ..., X_m, \tilde{c}_2^{(j)}, \tilde{c}_3^{(j)}, ..., \tilde{c}_{n-m+1}^{(j)})^T, \quad (14)$$

where

$$X_p = \sum_{l=1}^{p} \frac{(-it)^l}{(l-1)!} b_{p-l+1}^{(j)} e^{-i\omega_1^{(j)}t}, \quad \tilde{c}_l^{(j)} = c_l^{(j)} e^{-i\omega_l^{(j)}t}, \quad (15)$$

with $b_{p-l+1}^{(j)}$ and $c_l^{(j)}$ the coefficients to be determined according to the boundary conditions. Then, the wavefunction in original basis is given by

$$\vec{\psi}\,' = \hat{M}_j^{-1} \cdot \vec{\psi} = \sum_{p=1}^{m} X_p \vec{\psi}_{1,p}^{(j)} + \sum_{l=2}^{n-m+1} \tilde{c}_l^{(j)} \vec{\psi}_l^{(j)}. \quad (16)$$

According to Eqs. (15) and (16), the time-evolution behavior of the wavefunction under a defective Hamiltonian $\hat{H}_j$ differs significantly from that of a diagonalizable one. In sharp contrast to the diagonalizable case—where the fields either decay or grow exponentially for complex eigenvalues—certain field components in the defective case diverge following a power-law dependence on time. Consequently, the wavefunction experiences variations in both phase and amplitude, making the phase delay matrix alone insufficient to capture the full time evolution within a homogeneous temporal layer governed by a defective Hamiltonian. This necessitates the introduction of a new class of transfer matrices.

Similarly, we denote $\vec{\beta}_j = (b_1^{(j)}, b_2^{(j)}, ..., b_m^{(j)}, c_2^{(j)}, ..., c_{n-m+1}^{(j)})^T$, characterizing the expansion coefficients of the wavefunction in the canonical basis at the beginning of j-th temporal layer, namely $\vec{\beta}_j = \vec{\psi}\,'(t=0)$. As such, according to Eq. (15), we obtain

$$\vec{\psi}(t) = \hat{M}_j \cdot \vec{\psi}\,'(t) = \hat{M}_j \cdot \hat{B}_j(t) \cdot \hat{D}_j(t) \cdot \vec{\beta}_j, \quad (17)$$

where

$$\hat{D}_j(t) = (e^{-i\omega_1^{(j)}t} \hat{I}_m) \oplus \mathrm{diag}\{e^{-i\omega_2^{(j)}t}, e^{-i\omega_3^{(j)}t}, ..., e^{-i\omega_{n-m+1}^{(j)}t}\}, \quad (18)$$

is the new phase delay matrix within the j-th temporal layer, and

$$\hat{B}_j(t) = \hat{A}_m(t) \oplus \hat{I}_{n-m}, \quad (19)$$

with

$$\hat{A}_m(t) = \begin{pmatrix} 1 & & & \\ -it & 1 & & \\ \vdots & \ddots & \ddots & \\ \frac{(-it)^m}{(m-1)!} & \cdots & -it & 1 \end{pmatrix}. \quad (20)$$

Consequently, besides the temporal phase delay matrix $\hat{D}_j(t)$, an additional matrix $\hat{B}_j(t)$ arises, representing the amplitude boosting. Thus, we call $\hat{B}_j(t)$ as the amplitude-boosting matrix within the j-th temporal layer. From Eq. (20), the amplitude boosting matrix is irrelevant to the eigenvalues or the generalized eigenvectors, but depends solely on the order of the EP.

According to the temporal boundary conditions that $\vec{\psi}(t)$ is continuous, the temporal matching matrix $\hat{M}_{j,j+1}$ is also defined by Eq. (9). Thus, $\vec{\beta}_j$ is transferred to $\vec{\beta}_{j+1}$ according to

$$\vec{\beta}_{j+1} = \hat{M}_{j,j+1} \cdot \hat{B}_j(t_j) \cdot \hat{D}_j(t_j) \cdot \vec{\beta}_j. \quad (21)$$

When no EP is present, corresponding to $m=1$, $\hat{A}_m(t)$ reduces to 1 or equivalently $\hat{I}_1$. Consequently, the amplitude-boosting matrix $\hat{B}_j(t)$ becomes $\hat{I}_1 \oplus \hat{I}_{n-1} = \hat{I}_n$ and can be omitted from the matrix product, reducing the TTMM to that for a diagonalizable Hamiltonian.

Furthermore, when multiple EPs of different orders are present, the canonical basis can be constructed in the same manner, leading to the Jordan normal form $\hat{J}_j$ composed of several Jordan blocks. In this case, the corresponding phase-delay and amplitude-boosting matrices can be defined analogously:

$$\hat{D}_j(t) = (e^{-i\omega_1^{(j)}t}\hat{I}_{m_1}) \oplus (e^{-i\omega_2^{(j)}t}\hat{I}_{m_2}) \oplus \cdots \oplus (e^{-i\omega_p^{(j)}t}\hat{I}_{m_p}), \quad (22)$$

and

$$\hat{B}_j(t) = \hat{A}_{m_1}(t) \oplus \hat{A}_{m_2}(t) \oplus \cdots \oplus \hat{A}_{m_p}(t), \quad (23)$$

where $m_1 + m_2 + \cdots + m_p = n$. $m_l > 1$ denotes the order of the EP. For $m_l = 1$, $\hat{A}_{m_l}(t) = \hat{I}_1$, corresponding to the non-EP case.

### III. Drude Media

In this section, we present a typical system that exhibits defective Hamiltonians—namely, lossless Drude media. Within a generic Drude medium, the polarization charge $P_x$ and electric field $E_x$ are governed by

$$\partial_{tt}P_x + \gamma\partial_t P_x = \omega_p^2 E_x, \quad (24)$$

where $\omega_p$ and $\gamma$ denote the plasma frequency and damping rate, respectively. The vacuum permittivity $\varepsilon_0$ and permeability $\mu_0$ are set as unities for simplicity. When the magnetic field is directed along the y-axis and the wavenumber vector lies along the z-axis, Maxwell's equations are expressed as

$$\partial_z E_x = -\partial_t H_y, \quad \partial_z H_y = -\partial_t E_x - \partial_t P_x. \quad (25)$$

Combining Eqs. (24) and (25), we obtain the Hamiltonian and wavefunction as

$$\hat{H}^{(Drude)} = \begin{pmatrix} 0 & -i\partial_z & 0 & -i \\ -i\partial_z & 0 & 0 & 0 \\ 0 & 0 & 0 & i \\ i\omega_p^2 & 0 & 0 & -i\gamma \end{pmatrix}, \quad \vec{\psi} = \begin{pmatrix} E_x \\ H_y \\ P_x \\ J_x \end{pmatrix}. \quad (26)$$

For plane wave modes with a spatial dependence $e^{ikz}$, where $k$ is the wavenumber, $-i\partial_z$ can be replaced by $k$ in $\hat{H}^{(Drude)}$. Eq. (26) shows that $\hat{H}^{(Drude)}$ turns defective in the lossless limit ($\gamma = 0$), leading to an exceptional line (EL) of order 2 pinned at zero frequency in the dispersion relation $\omega(k)$ [24]. This EL is typically overlooked in time-invariant systems, as it cannot be excited when the incident frequency is finite. In contrast, in time-varying systems, the frequency is no longer conserved, making the EL accessible and observable.

When $\gamma = 0$, the eigenvalues and corresponding ordinary eigenvectors of $\hat{H}^{(Drude)}$ are given by:

$$\omega_{1-3} = \{0, -\sqrt{k^2 + \omega_p^2}, \sqrt{k^2 + \omega_p^2}\}, \quad (27)$$

and

$$(\vec{\psi}_1, \vec{\psi}_2, \vec{\psi}_3) = \begin{pmatrix} 0 & -i\omega_2 \omega_p^{-2} & -i\omega_3 \omega_p^{-2} \\ 0 & -ik\omega_p^{-2} & -ik\omega_p^{-2} \\ 1 & i\omega_2^{-1} & i\omega_3^{-1} \\ 0 & 1 & 1 \end{pmatrix} e^{ikz}. \quad (28)$$

Using the Jordan chain Eq. (10), the generalized eigenvector of rank 2 corresponding to the eigenvalue $\omega_1$ is obtained as:

$$\vec{\psi}_{1,2} = (0, k^{-1}, 0, -i)^T e^{ikz}. \quad (29)$$

The third element of $\vec{\psi}_{1,2}$ can be arbitrary; however, for simplicity, we set it to zero here. Therefore, the ordinary eigenvectors in Eq. (28) and the generalized eigenvector in Eq. (29) construct the canonical basis of $\hat{H}^{(Drude)}$ in the lossless limit $\gamma = 0$. Based on the expansion in the canonical basis, the generalized modal, phase-delay and amplitude-boosting matrices are given by

$$\hat{M} = (\vec{\psi}_{1,2}, \vec{\psi}_1, \vec{\psi}_2, \vec{\psi}_3), (30)$$

$$\hat{D}(t) = \text{diag}\{e^{-i\omega_1 t}, e^{-i\omega_1 t}, e^{-i\omega_2 t}, e^{-i\omega_3 t}\}, (31)$$

and

$$\hat{B}(t) = \hat{A}_2(t) \oplus \hat{I}_2. (32)$$

## A. Multilayer with alternating plasma frequencies

The time-evolution of electromagnetic waves in temporal multilayer with alternating plasma frequencies can be determined through the successive multiplication of the corresponding transfer matrices. The plasma frequency the j-th temporal layer is indicated by $\omega_{pj}$. Then, inserting $\omega_p = \omega_{pj}$ into Eqs. (27)-(32), the generalized modal $\hat{M}_j$, phase-delay $\hat{D}_j(t)$ and amplitude-boosting $\hat{B}_j(t)$ matrices are obtained. The beginning and ending time moments of the j-th temporal layer are indicated by $t_{j-1}$ and $t_j$, respectively. Suppose that the initial wavefunction is given by $\vec{\psi}(t_0)$, the vector for the expansion coefficients in the canonical basis is expressed as

$$\vec{\beta}_1 = \hat{M}_1^{-1} \cdot \vec{\psi}(t_0). (33)$$

Then, within the first temporal layer, the wavefunction is calculated as

$$\vec{\psi}(t_0 < t \leq t_1) = \hat{M}_1 \cdot \hat{B}_1(t - t_0) \cdot \hat{D}_1(t - t_0) \cdot \vec{\beta}_1. (34)$$

For the second temporal layer, $\vec{\beta}_2$ is obtained by imposing the matching matrix:

$$\vec{\beta}_2 = \hat{M}_{1,2} \cdot \hat{B}_1(t_1 - t_0) \cdot \hat{D}_1(t_1 - t_0) \cdot \vec{\beta}_1, (35)$$

and the wavefunction is

$$\vec{\psi}(t_1 < t \leq t_2) = \hat{M}_2 \cdot \hat{B}_2(t - t_1) \cdot \hat{D}_2(t - t_1) \cdot \vec{\beta}_2. (36)$$

Likewise, $\vec{\beta}_{j+1}$ and the wavefunction $\vec{\psi}(t)$ in the (j+1)-th temporal layer can be obtained iteratively according to

$$\vec{\beta}_{j+1} = \hat{M}_{j,j+1} \cdot \hat{B}_j(t_j - t_{j-1}) \cdot \hat{D}_j(t_j - t_{j-1}) \cdot \vec{\beta}_j, (37)$$

and

$$\vec{\psi}(t)(t_j < t \leq t_{j+1}) = \hat{M}_{j+1} \cdot \hat{B}_{j+1}(t - t_j) \cdot \hat{D}_{j+1}(t - t_j) \cdot \vec{\beta}_{j+1}. (38)$$

Figure 1 presents the field components of a plane wave propagating through two temporal layers of lossless Drude media with $\omega_{p2} = 2\omega_{p1}$. The circles denote TTMM results obtained from Eqs. (33)–(38). The polarization $P_x$ (blue curves) exhibits temporal growth, revealing the amplitude-boosting effect. However, since no gain medium is introduced, the temporal growth of the polarization charge does not imply an increase in the total electromagnetic energy [24].

The time-evolution of the wavefunction can also be obtained by solving Eq. (1) directly, yielding

$$\vec{\psi}(t) = \mathcal{T}\left[\exp[-i\int_{t_0}^{t} \hat{H}(t')dt']\right] \cdot \vec{\psi}(t_0), (39)$$

where $\mathcal{T}$ denotes the time-ordering operator, and $\hat{H}(t)$ is the time-dependent Hamiltonian that takes $\hat{H}_j^{(Drude)}$ in the j-th temporal layer. The solid lines in Figure 1 represent results from Eq. (39). The excellent agreement between the circles and solid lines verifies the accuracy and validity of the proposed TTMM.

## B. Floquet band structures

The medium becomes a PTC when the plasma frequency is time-periodic. Consider that the unit cell consists of two temporal sublayers, A and B. Both sublayers are lossless, but they have distinct plasma frequencies, $\omega_{pa}$ and $\omega_{pb}$. The vector for expansion coefficients in the canonical basis at the beginning of sublayer A (B) is indicated as $\vec{\beta}_a$ ($\vec{\beta}_b$). Based on the Floquet theorem, we obtain

$$\hat{M}_{b,a} \cdot \hat{B}_b(t_b) \cdot \hat{D}_b(t_b) \cdot \hat{M}_{a,b} \cdot \hat{B}_a(t_a) \cdot \hat{D}_a(t_a) \cdot \vec{\beta}_a = \vec{\beta}_a e^{-iQT}, (40)$$

where $t_a, t_b$ are the durations of the sublayers, $T = t_a + t_b$ is the modulation period, $Q$ denotes the quasienergy, $\hat{D}_a, \hat{D}_b$ represent the phase-delay matrices, $\hat{M}_{b,a} = \hat{M}_a^{-1} \cdot \hat{M}_b = \hat{M}_{a,b}^{-1}$ are the temporal matching matrices. Eq. (40) also holds when the subscripts a and b are interchanged, and it can be further extended to a PTC whose unit cell contains an arbitrary number of sublayers. Nontrivial solutions of Eq. (40), corresponding to the eigenmodes, exist when the following secular equation is satisfied:

$$\det | \hat{M}_{b,a} \cdot \hat{B}_b(t_b) \cdot \hat{D}_b(t_b) \cdot \hat{M}_{a,b} \cdot \hat{B}_a(t_a) \cdot \hat{D}_a(t_a) - e^{-iQT} | = 0. \quad (41)$$

To verify the validity of Eq. (41), we introduce the Floquet matrix method, where the quasienergy bands can be obtained as the eigenvalues of the following Flqouet matrix [34]:

$$\hat{H}_F = \begin{pmatrix} \ddots & & & & \\ & \hat{H}_0 + \Omega & \hat{V}_1 & \hat{V}_2 & \\ & \hat{V}_{-1} & \hat{H}_0 & \hat{V} & \\ & \hat{V}_{-2} & \hat{V}_{-1} & \hat{H}_0 - \Omega & \\ & & & & \ddots \end{pmatrix}, \quad (42)$$

where $\Omega = 2\pi/T$ is the modulation frequency, and

$$\hat{H}_0 = \frac{1}{T} \int_0^T \hat{H}(t) dt, \quad \hat{V}_j = \frac{1}{T} \int_0^T \hat{H}(t) e^{-ij\Omega t} dt, \quad (43)$$

are the zeroth and nonzeroth order Fourier series of $\hat{H}(t)$. For the temporal AB lattice, one has

$$\hat{H}_0 = \begin{pmatrix} 0 & -i\partial_z & 0 & -i \\ -i\partial_z & 0 & 0 & 0 \\ 0 & 0 & 0 & i \\ i\omega_{p0}^2 & 0 & 0 & 0 \end{pmatrix}, \quad (44)$$

where $\omega_{p0}^2 = (\omega_{pa}^2 t_a + \omega_{pb}^2 t_b)/T$, and in $\hat{V}_j$, all matrix elements vanish except for the entry at the 4th row and 1st column:

$$V_{j,41} = \frac{i(\omega_{pa}^2 - \omega_{pb}^2)}{j\pi} \sin\left(\frac{j\pi t_a}{T}\right). \quad (45)$$

Figure 2 shows the band structure of a typical PTC within the first Floquet zone. Both the TTMM [Eq. (41)] and the Floquet matrix method [Eq. (42)] are employed, with the corresponding results shown as red circles and black solid lines, respectively. The two sets of results coincide excellently in both their real and imaginary parts, further confirming the accuracy and validity of the proposed TTMM.

### C. Transition to weakly lossy cases

When $\gamma$ is nonzero, $\hat{H}^{(Drude)}$ becomes diagonalizable, and the EL disappears. Formally, in this case, the modal matrices are constructed from four linearly independent ordinary eigenvectors of $\hat{H}^{(Drude)}$, the amplitude-boosting matrices are omitted, and the phase-delay matrices are determined by the four distinct eigenvalues. Nevertheless, for sufficiently small $\gamma$, the transfer matrices defined in Eqs. (30)–(32), which are for the EP case, can still accurately capture the time evolution of electromagnetic waves over a considerably long duration.

For demonstration, Figures 3 (a)–3(c) show the real parts of the electric, magnetic, and polarization-charge fields for a plane wave propagating inside a PTC composed of weakly lossy Drude dispersive layers. The solid lines correspond to results from the rigorous calculation, while the circles represent results obtained under the lossless approximation, in which the amplitude-boosting matrices are included.

For $\gamma = 0.01\omega_p$, as shown in Figures 3(a) and 3(b), the electric and magnetic fields are accurately predicted over five unit cells under the lossless approximation. However, a visible discrepancy appears in $P_x$ after three unit cells, as shown in Figure 3(c). This is mainly because only the amplitude of $P_x$ undergoes temporal boosting. Under the lossless approximation, the amplitude-boosting matrices predict a power-law growth in time, which deviates from the actual exponential growth, especially after a long time. However, the discrepancy is reduced for smaller $\gamma$, as indicated by the green and red lines in Figure 3(c). In principle, when $\gamma$ is sufficiently small, the discrepancy can be ignored over a sufficient long time.

### IV. Lorentz dispersive media with extremely high loss

We present another representative system — the Lorentz dispersive medium — which can exhibit higher-order EPs in its band dispersion when significant loss is introduced [23]. The polarization charge in a Lorentz dispersive medium is governed by

$$\partial_{tt} P_x + \gamma \partial_t P_x + \omega_0^2 P_x = \omega_p^2 E_x, \quad (46)$$

where $\omega_0$ is the resonant frequency. While the wavefunction $\vec{\psi}$ is still defined in Eq. (26), the Hamiltonian for the Lorentz dispersive medium is transformed to

$$\hat{H}^{(Lorentz)} = \begin{pmatrix} 0 & -i\partial_z & 0 & -i \\ -i\partial_z & 0 & 0 & 0 \\ 0 & 0 & 0 & i \\ i\omega_p^2 & i\omega_0^2 & 0 & -i\gamma \end{pmatrix}. \quad (47)$$

Our previous work demonstrated that two ELs appear in the parameter space ($\omega, \gamma$) when $k = \omega_0 / c$ and $\omega_p = 0.5\gamma$, and they merge to form an EP-4 at $\gamma = 4\omega_0$ [23].

We focus on the EP-4, where the eigenvalue is fourfold degenerate at $\omega = -i\omega_0$. Based on the Jordan chain in Eq. (10), the corresponding generalized eigenvectors are obtained as

$$\vec{\psi}_{1,1} = (\frac{1}{2\omega_0}, \frac{i}{2\omega_0}, -\frac{1}{\omega_0}, 1)^T e^{ikz}, \quad (48)$$

$$\vec{\psi}_{1,2} = (\frac{1}{2\omega_0}, \frac{1+i\omega_0}{2\omega_0^2}, \frac{i-\omega_0}{\omega_0^2}, 1)^T e^{ikz}, \quad (49)$$

$$\vec{\psi}_{1,3} = (\frac{1+2\omega_0^2}{4\omega_0^3}, \frac{-i+2\omega_0+2i\omega_0^2}{4\omega_0^3}, \frac{1+i\omega_0-\omega_0^2}{\omega_0^3}, 1)^T e^{ikz}, \quad (50)$$

and

$$\vec{\psi}_{1,4} = (\frac{-i+\omega_0+2\omega_0^3}{4\omega_0^4}, \frac{-i+2\omega_0+2i\omega_0^2}{4\omega_0^3}, \frac{-i+\omega_0+i\omega_0^2-\omega_0^3}{\omega_0^4}, 1)^T e^{ikz}. \quad (51)$$

They form the columns of the generalized modal matrix of $\hat{H}^{(Lorentz)}$. The phase-delay and amplitude-boosting matrices are simply expressed as

$$\hat{D}(t) = e^{-\omega_0 t} \hat{I}_4, \quad \hat{B}(t) = \hat{A}_4(t). \quad (52)$$

To verify the validity of Eqs. (48)–(52), we show the electromagnetic fields of a plane wave propagating through three temporal layers, in which a lossless Drude medium is sandwiched between lossy Lorentz dispersive media along the time axis. The parameters of the Lorentz dispersive media are chosen to realize the EP-4. Accordingly, when applying the TTMM, Eqs. (48)–(52) are used for the Lorentz dispersive layers, while Eqs. (27)–(32) are applied for the Drude medium. As seen in Figure 4, the TTMM results (circles) coincide perfectly with the direct calculations from Eq. (39) (solid lines), demonstrating the excellent accuracy and reliability of the TTMM for systems involving EPs of different orders.

V. **Conclusions**

To summarize, we have developed the TTMM by expanding the wavefunctions in the canonical basis of the Hamiltonian. In particular, the TTMM has been extended to exceptional-point (EP) media systems described by defective Hamiltonians. In addition to the temporal phase-delay and matching matrices, amplitude-boosting matrices are introduced to characterize the power-law growth of field amplitudes in time. The applicability of the developed TTMM is demonstrated in two representative EP media systems: the lossless Drude medium exhibiting an EP-2 and the lossy Lorentz dispersive medium exhibiting an EP-4. The numerical results confirm that the TTMM provides an accurate and efficient framework for evaluating electromagnetic fields and Floquet band dispersions.

**Acknowledgements:** The Key Project of the National Key R&D program of China (No. 2022YFA1404500), and National Natural Science Foundation of China (NSFC) (No. 12174263 and No. 12074267)

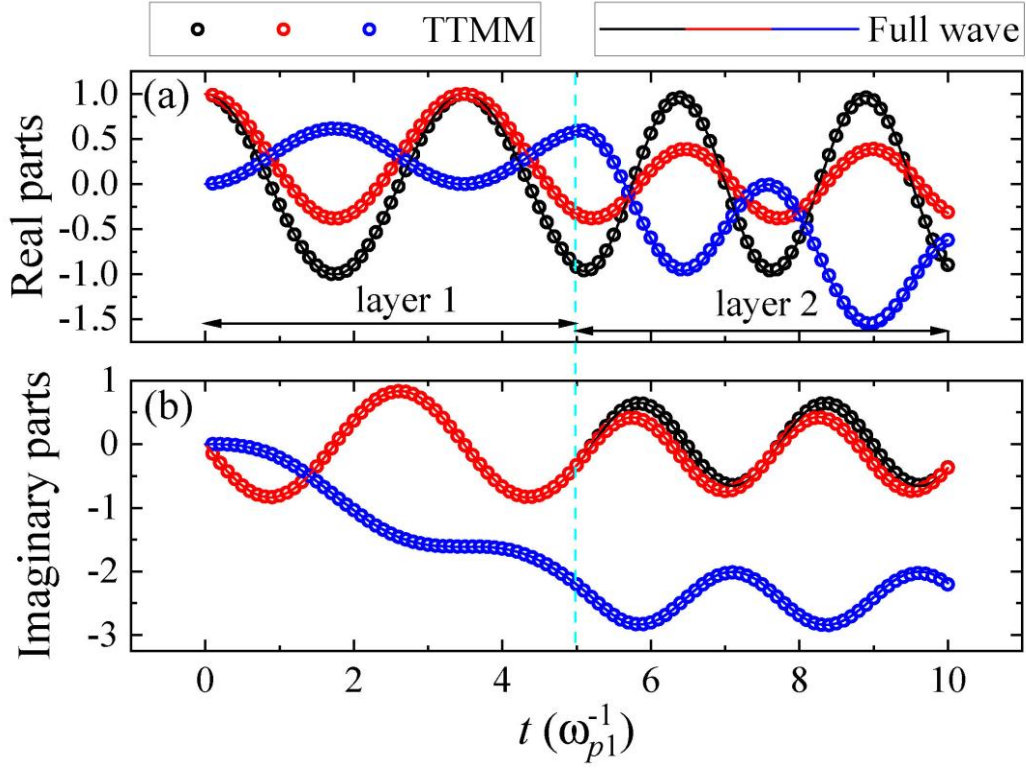

Figure 1. (a) Real and (b) imaginary parts of the electromagnetic fields obtained from full-wave simulations using Eq. (39) (solid lines) and from the TTMM based on Eq. (38) (circles). The field components $E_x$, $H_y$, and $P_x$ are represented by black, red, and blue curves, respectively. The initial wave function is $\vec{\psi}(t_0) = (1,1,0,0)^T e^{ikz}$, where $k = 1.5\omega_{p1}/c$, and the plasma frequencies of the two temporal layers (separated by the cyan dashed line) are $\omega_{p1}$ and $\omega_{p2} = 2\omega_{p1}$.

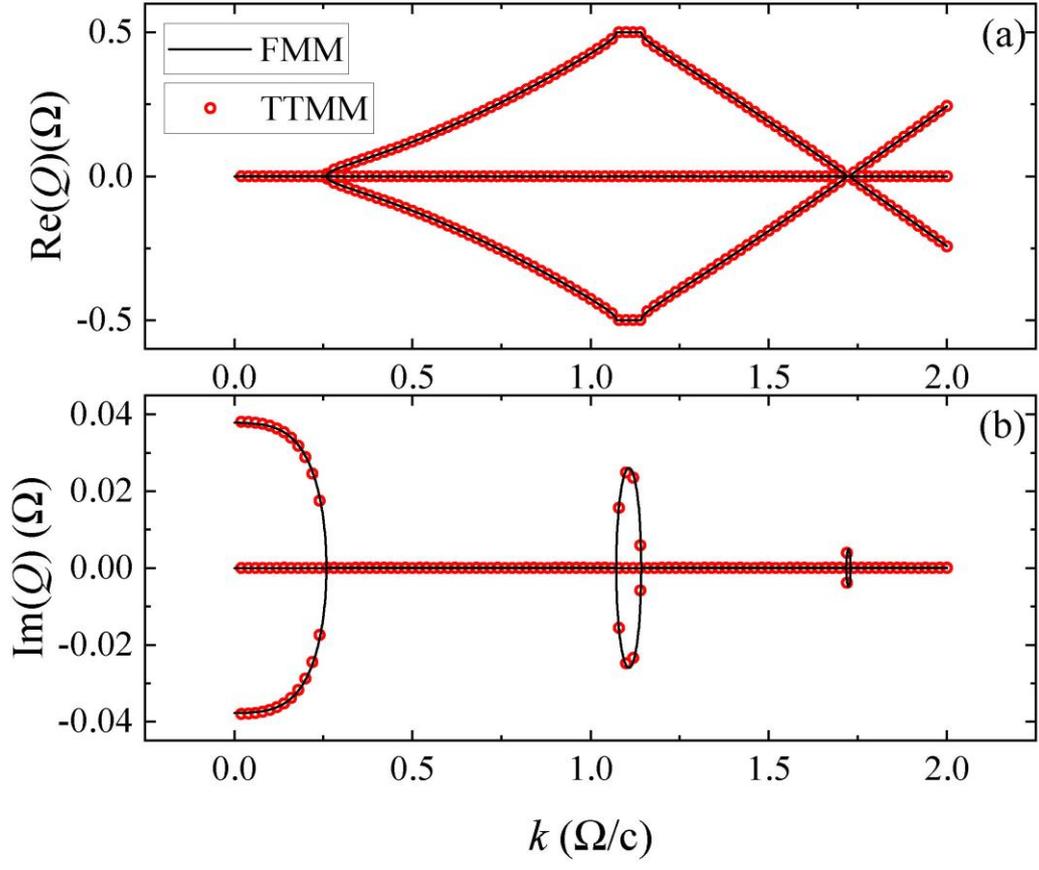

Figure 2. (a) Real and (b) imaginary parts of quasienergy bands in the first Floquet zone ( $\mathrm{Re}(Q) \in [-\Omega/2, \Omega/2]$ ). The results obtained by the Flqouet matrix method (FMM) and the TTMM are shown by the black lines and red circles, respectively. The PTC is composed of two alternating sublayers, with plasma frequencies and durations: $\omega_{pa} = 0.8\Omega$, $\omega_{pb} = 1.2\Omega$ and $t_a = t_b = 0.5T$.

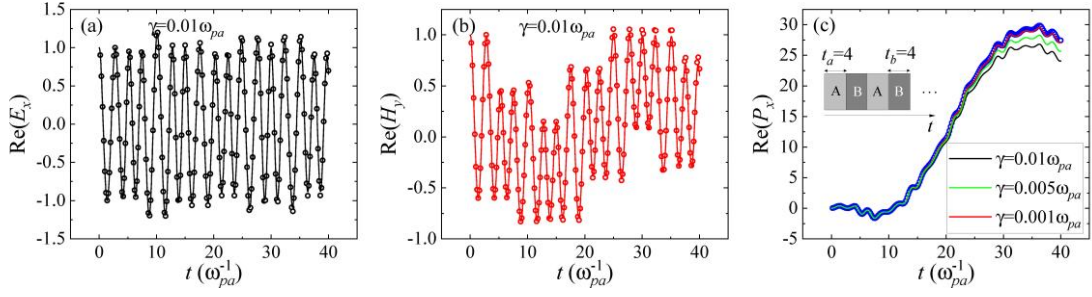

Figure 3. (a) $\text{Re}(E_x)$, (b) $\text{Re}(H_y)$, and (c) $\text{Re}(P_x)$, for an electromagnetic wave propagating in a PTC. The solid lines represent results obtained using the TTMM based on diagonalizable Hamiltonians ($\gamma \neq 0$), while the circles correspond to those obtained using the TTMM based on the defective Hamiltonian ($\gamma = 0$). The parameters of the two temporal sublayers are $t_a = t_b = 4/\omega_{pa}$ and $\omega_{pb} = 2\omega_{pa}$, and the initial wave function $\vec{\psi}(t_0) = (1,1,0,0)^T e^{ikz}$ with $k = 2\omega_{pa}/c$. In panels (a) and (b), $\gamma = 0.01\omega_{pa}$, while in panel (c), $\gamma = 0.01\omega_{pa}$ (black), $0.005\omega_{pa}$ (green) and $0.001\omega_{pa}$ (red) are considered.

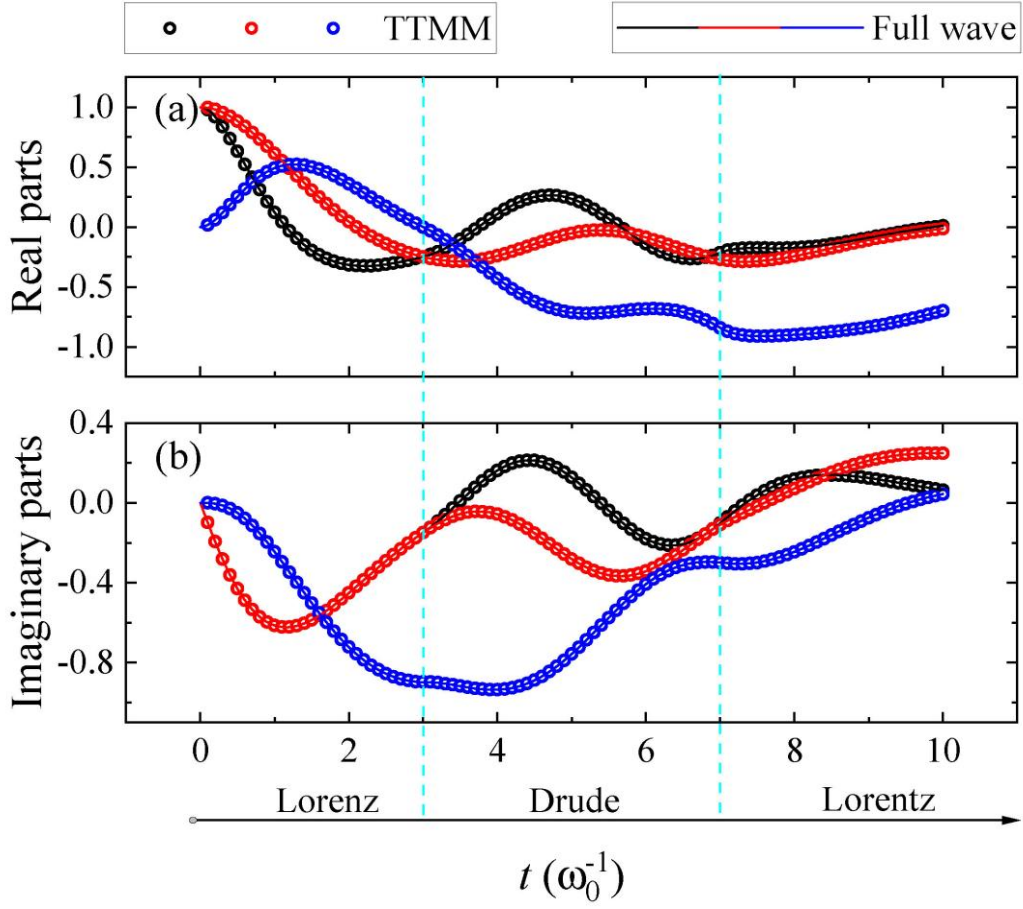

Figure 4. (a) Real and (b) imaginary parts of the electromagnetic fields obtained from full-wave simulations (solid lines) and from the TTMM circles). The field components $E_x$, $H_y$, and $P_x$ are represented by black, red, and blue curves, respectively. The cyan dashed lines mark the photonic time interfaces. The medium is initially Lorentz-dispersive, exhibiting an EP-4, then transitions into a lossless Drude-type medium exhibiting an EP-2, and finally returns to the Lorentz-dispersive state. For the Lorentz dispersion, $\omega_0 = kc$, $\omega_p = 2\omega_0$, $\gamma = 4\omega_0$, while for the Durde dispersion, $\omega_p = 1.3\omega_0$, $\gamma = 0$. The wavefunction is initially $\vec{\psi}(t_0) = (1,1,0,0)^T e^{ikz}$.